\documentclass[namedreferences,hyperref,optionalrh,solaromanenum]{spr-sola}

\usepackage{graphicx}                    % For eps figures, newer & more powerfull
\usepackage{color}                       % For color text: \color command
\chardef\us=`\_

\begin{document}

\begin{frontmatter}

\title{Hydrogen Ionization Inside the Sun}

%%%%%%%%%%%%%%%%%%%%%%%%%%%%%%%%%%%%%%%%%%%%%%%%%%%
%% Authors Names
%
% \author[addressref={},corref,email={}]{\inits{}\fnm{}\snm{}\orcid{}}
\author[addressref={1}]{\inits{V.A.}\fnm{Vladimir A. }\snm{Baturin}\orcid{0009-0000-7642-6786}}
\author[addressref={1}]{\inits{S.V. }\fnm{Sergey V. }\snm{Ayukov}\orcid{0009-0001-8859-3570}}
\author[addressref={1},corref,email={avo@sai.msu.ru}]{\inits{A.V.}\fnm{Anna V. }\snm{Oreshina}\orcid{0009-0004-7969-1840}}
\author[addressref={1}]{\inits{A.B. }\fnm{Alexey B. }\snm{Gorshkov}}
\author[addressref={2}]{\inits{V.K. }\fnm{Victor K. }\snm{Gryaznov}\orcid{0000-0003-4167-5090}}
\author[addressref={3,4}]{\inits{I.L. }\fnm{Igor L. }\snm{Iosilevskiy}}
\author[addressref={5}]{\inits{W. }\fnm{Werner }\snm{D\"appen}}
%%%%%%%%%%%%%%%%%%%%%%%%%%%%%%%%%%%%%%%%%%%%%%%%%%%
%% Runningheads
%
\runningauthor{V.A. Baturin et al.}
\runningtitle{Hydrogen ionization inside the Sun}

%%%%%%%%%%%%%%%%%%%%%%%%%%%%%%%%%%%%%%%%%%%%%%%%%%%
%% Affilations 
%% id shold be the same with \author addressref value.
%\address[id={}]{}
\address[id={1}]{Sternberg Astronomical Institute, M.V. Lomonosov Moscow State University, Moscow, Russia}
\address[id={2}]{Federal Research Center of Problems of Chemical Physics and Medicinal Chemistry RAS, Chernogolovka, Russia}
\address[id={3}]{Joint Institute for High Temperatures RAS, Moscow, Russia}
\address[id={4}]{Moscow Institute of Physics and Technology, Dolgoprudnyi, Russia}
\address[id={5}]{Department of Physics and Astronomy, University of Southern California, Los Angeles, CA 90089, USA}
%%%%%%%%%%%%%%%%%%%%%%%%%%%%%%%%%%%%%%%%%%%%%%%%%%%
%%% Abstract 
\begin{abstract}
Hydrogen is the main chemical component of the solar plasma, and H-ionization determines basic properties of the first adiabatic exponent ${\Gamma _1}$. Hydrogen ionization remarkably differs from the ionization of other chemicals. Due to the large number concentration, H-ionization causes a very deep lowering of ${\Gamma _1}$, and the lowering profile appears to be strongly asymmetric and extends over almost the entire solar convective zone. 
The excited states in the hydrogen atom are modelled with the help of a partition function, which accounts the internal degrees of freedom of the composed particle. A temperature-dependent partition function with an asymptotic cut-off tail is deduced from a solution of the quantum mechanical problem of the hydrogen atom in the plasma. We present a numerical simulation of hydrogen ionization, calculated with two expressions for the partition function, Planck-Larkin (PL) and Starostin-Roerich (SR), respectively. The Hydrogen ionization is shifted toward higher temperature in the SR-case compared to the PL-case. 
Different models for excited states of the hydrogen atom may change ${\Gamma _1}$ by as much as $10^{-2}$. The behavior of the ${\Gamma _1}$  profiles for pure hydrogen resembles “twisted ropes” for the two considered models. This significantly affects the helium ionization and the position of the helium hump. This  entanglement of H and He effect gives us a chance to study a role of excited states in the solar plasma.
	
\end{abstract}

%%%%%%%%%%%%%%%%%%%%%%%%%%%%%%%%%%%%%%%%%%%%%%%%%%%
%% Keywords
%
\end{frontmatter}
%-------------------------------------------------

%%%%%%%%%%%%%%%%%%%%%%%%%%%%%%%%%%%%%%%%%%%%%%%%%%%
%% Sections
%
 \section{Introduction}%\label{s:?} 
 
The equation of state (EOS) is a key element in the modeling of the solar structure. Its role is especially obvious in the problem of helioseismic inverse modeling to restore physical features of the solar plasma. The equation of state is used to compute measurable thermodynamic functions (pressure, adiabatic exponent, specific heat capacity, etc.) for given temperature, density, and chemical composition of equilibrium plasma.

\begin{figure} 
\centerline{\includegraphics[width=1.0\textwidth,clip=]{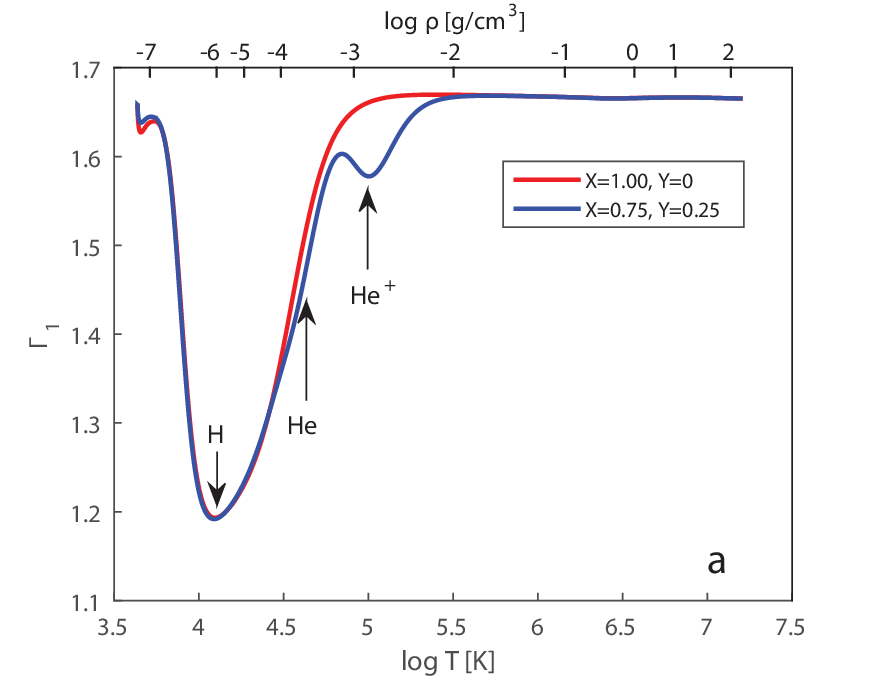}}
\centerline{\includegraphics[width=1.0\textwidth,clip=]{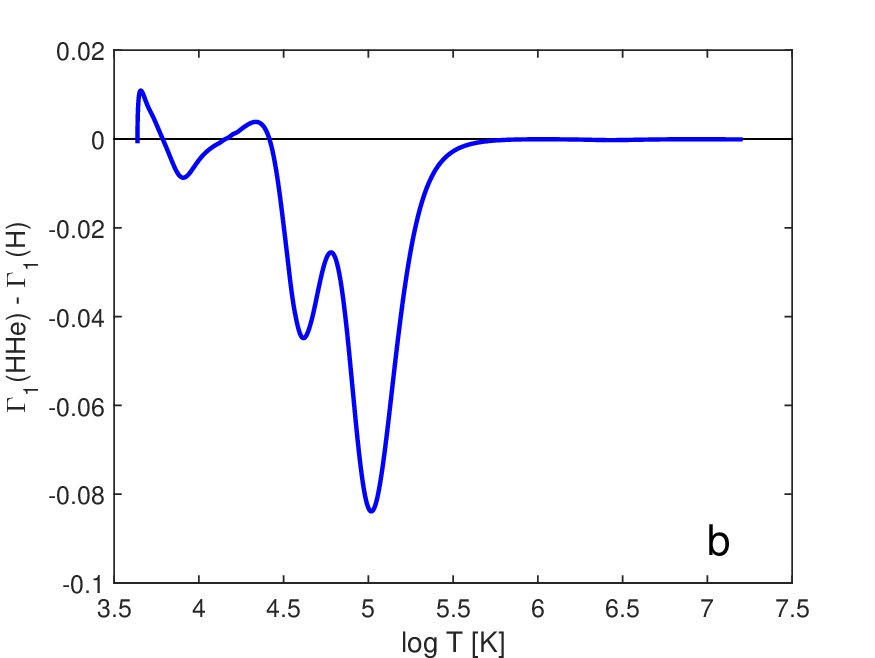}}
\caption{(a) Adiabatic exponent profiles for hydrogen (\textit{red curve}) and hydrogen-helium (\textit{blue curve}) plasmas calculated at points $(T,\rho)$ from the standard solar model. (b) Difference between ${\Gamma _1}$ of hydrogen-helium and hydrogen plasma.} 
\label{Fig_G1_HHe}
\end{figure}

The most striking and obvious physical effect in the theory of the EOS under solar conditions is the ionization of elements, among which hydrogen and helium give the strongest effect both in terms of pressure $P$ and adiabatic exponent $\Gamma_1$, describing elasticity of the plasma. $\Gamma_1$ is variation of pressure  under adiabatic compression: ${\Gamma _1}=\partial \ln P /\partial \ln\rho |_S$. We consider how ionization affects the adiabatic exponent, calculated along some fixed sequence of points $\left( {T,\rho } \right)$ of a standard solar model. For illustration, examples of  ${\Gamma _1}$ profiles are shown in Figure~\ref{Fig_G1_HHe}. The red curve shows the ${\Gamma _1}$ profile for a purely hydrogen plasma, while the blue curve shows the ${\Gamma _1}$ profile for a plasma composed of hydrogen and helium, with mass contents of hydrogen $X = 0.75$ and helium $Y = 0.25$, which is close to the solar content. 

%%%
%%%

The ionization of hydrogen is strikingly different from that of other elements. First, due to the abundance of hydrogen (90\% in number of particles), the decrease of ${\Gamma _1}$ in the ionization region turns out to be very strong and the value of  ${\Gamma_1}$ falls from values around 5/3 to 1.2. Secondly, the ionization of hydrogen occurs against the background of an almost complete absence of free electrons from other ions. Therefore, mainly electrons from the ionization of hydrogen itself must be taken into account. This leads to the fact that the initial ionization of hydrogen occurs very abruptly and at very low temperatures compared to the ionization energy, i.e. ${I_H}/kT > 10$. Here ${I_H}$ is the ionization energy of the hydrogen atom from the ground state, and $k$ is the Boltzmann constant. The decrease of  ${\Gamma _1}$ occurs almost by a jump, whereas the subsequent transition of the curve to full ionization extends over a significant area in depth, up to temperatures ${I_H}/kT < 1$, which extends over a significant part of the convective zone

Thus, an accurate calculation of the ${\Gamma_1}$ profile in the hydrogen ionization region is a difficult task.
The question of the equation of state of hydrogen in a wide range of conditions has been considered in many works in the physical literature. The most advanced approaches were considered by \cite{Militzer_Ceperley_2001,Militzer_Hubbard_2013,  Wendland_2014, Chabrier_2019, Chabrier_2021, Filinov_2023}.
But we do not know application of these results to solar modeling. 
In the astrophysical application for modeling stars, the following EOS calculations are known:
\cite{Mihalas_1988, Rogers_1996, Rogers_Nayfonov_2002, Irwin_2012}. Appendix~\ref{Appendix_EOS} presents comparison of $\Gamma_1$ for available equations of state.

The main problem is related to the modeling the quantum structure of the hydrogen atom under conditions of sufficiently large density and high plasma temperatures. In other words, while the isolated hydrogen atom is a well-studied problem with known quantum states and statistical weights, in a plasma, the hydrogen atom is a difficult system due to the presence of nearby other particles with sufficiently high kinetic energy.  

The example of the ionization of pure hydrogen is quite far from the real conditions on the Sun. To approximate reality, we demonstrate also the role of helium ionization against the background of completed hydrogen ionization. Due to the larger first ionization potential, helium is ionized at higher temperatures than hydrogen, and the ratio to the ionization potential turns out to be ${I_{He}}/kT > 7$. The difficulty in ionization of helium is due to the fact that abundant free electrons from significantly ionized hydrogen are present. The lower panel of Figure 1 shows the difference in ${\Gamma_1}$ when helium is present compared to pure hydrogen. This difference serves as a qualitative estimate of the He contribution. Note that the He contribution generally resembles the theory of $Z$ contributions for He-like heavy element ions discussed in \citep{Baturin_2022}. It also follows from this figure that the first helium ionization leads to a lowering of the ${\Gamma_1}$ profile by about 0.05 at $\log T = 4.55$, while the second ionization lowers ${\Gamma_1}$ by about 0.09 at $\log T = 5$. 

The issue of theoretical modeling of the `ionization glitch' for helium ionization is discussed by \cite{Houdayer_2021}. 

We focus on solving a restricted problem of changing the profile of adiabatic exponent $\Gamma_1$ due to perturbation of the effective statistical weight of hydrogen in a narrow region inside the Sun, where transition from a neutral composition to a fully ionized state occurs. Hydrogen ionization occurs inside the Sun and in solar-type stars in the outermost envelope at relatively low temperature $T \sim 10^4 - 10^5~\mathrm{K}$. The density of the plasma is also quite low, $\rho \sim 10^{-7} - 10^{-3}~\mathrm{g\, cm^{-3}}$. 
	
The issues of hydrogen ionization beyond solar-like stars, for example in low-mass stars and planets, as well as the asymptotic behavior at high density and/or temperature regimes are outside the scope of this paper. 

The key object of our study is the so-called Planck-Larkin (PL) partition function. 
The history of the partition function goes back to one hundred years ago to work by \cite{Planck_2024}. This story is described in excellent details by W. Ebeling in historical essay \citep{Ebeling_2017}. 
Although the partition function formula was originally written by  \cite{Brillouin_1930} it is commonly referred to as the Planck-Larkin partition function under which name it was essentially rediscovered in the framework of “field-theoretical statistical thermodynamics by Vedenov and Larkin, 1960” (cited from \citealp{Ebeling_2017}; original papers are \citealp{Vedenov_Larkin_1959, Larkin_1960}). The further development of theory may be found in a series of papers by Ebeling, the most recent is \cite{Ebeling_2021}.  

We fix all the internal parameters of an EOS model, except of the partition function (PF hereafter). In particular, we keep the model of Coulomb interaction which is noticeable in the studied area. Moreover, there are 
many other assumptions: the degeneracy and relativism of electrons, the presence of molecules and negative ions and others. They are small and not discussed in the paper.

Our work was perfomed using SAHA-S EOS  (\citealp{Gryaznov_2004, Gryaznov_2006}, see also SAHA-S website \url{crydee.sai.msu.ru/SAHA-S_EOS}). SAHA-S EOS was created specifically for the analysis of thermodynamics at the conditions of the Sun and solar-type stars. It focuses  on those phenomena that are significant in the solar interior, leaving aside extreme domains of very high densities and/or very high temperatures. The basic thermodynamic quantities are computed with high numerical stability that provides smooth profiles of thermodynamic derivatives, including $\Gamma_1$. Some results on this issue can be found in \cite{Baturin_2019}. The numerical stability in calculating the pressure in SAHA-S is at the level of $10^{-8}$, and the stability of calculating derivatives is at the level of $10^{-6}$. This is important for the study of sophisticated physical effects and comparison between EOS models at the accuracy level necessary for helioseismic analysis.

%%%%%%%%%%%%%%%%%%%%%%%%%%%%%%%%%%%%%
\section{Ionization Modeling}
\label{Section_Ionization}

Ionization simulations are carried out within the framework of the chemical picture. The chemical picture is based on the minimization of the free energy $F\left( {T,V} \right)$ as a function of the concentrations of particles of different varieties: atoms, electrons, and protons in the case of hydrogen plasma at fixed $T$ and $V$  \citep{Harris_1960, Ebeling_1969, Graboske_1969}. These concentrations can be found from additional Saha's  ionization equilibrium equations \citep{Saha_1921}. The Saha equations are relations for the chemical potentials ${\mu _i}$ of the corresponding ions. For example, for the ionization of hydrogen $H \leftrightarrow p + {e^ - }$, the equation is written as
\begin{equation}
{\mu _H} - {\mu _p} - {\mu _e} = 0
\label{Eq_Saha_mu}
\end{equation}	

\noindent \citep{Prigogine_Defay_1954}. In the case of classical (nondegenerate, noninteracted) particles, from Equation~\ref{Eq_Saha_mu} the usual form of the Saha equation is obtained, which assumes the existence of the hydrogen atom only in the ground state:
\begin{equation}
\frac{n_p \, n_e}{n_H} = \frac{g_p \, g_e}{g_H} \mathchar'26\mkern-10mu\lambda _e^{-3}{e^{-I_H/kT}}, \quad
\mathchar'26\mkern-10mu\lambda _e = {\left( {\frac{2\pi {m_e}kT}{h^2}} \right)}^{1/2} .
\label{Eq_Saha}
\end{equation} 

\noindent
In this expression ${n_H}$, ${n_p}$, ${n_e}$ are concentrations of neutral hydrogen, protons, and free electrons respectively. The variables ${g_H}$, ${g_p}$, ${g_e}$ are the statistical weights of the particles, i.e. the parameters describing the internal degree of freedom of the particle. In the case under consideration, ${g_p} = 1$, ${g_e} = 2$, ${g_H} = 2$, while ${I_H}$ is the ionization potential for the ground state. The value ${\mathchar'26\mkern-10mu\lambda _e}$ defined through the classical constants is sometimes called the thermal wavelength of the electron. 

However, if we want to take into account the existence of hydrogen not only in the ground state, but also in the excited states, the equation of Saha (\ref{Eq_Saha}) can be rewritten as
\begin{equation}
\frac{{{n_H}}}{{n_e^{}n_p^{}}} = \mathchar'26\mkern-10mu\lambda _e^3 \cdot {e^{\beta {I_H}}}\sum\limits_{n = 1}^\infty  {{{\tilde g}_n}} {e^{ - \beta {\varepsilon _n}}}{w_n} \, .
\label{Eq_Saha_sum}
\end{equation} 

\noindent
In this case, the excited states are combined in the sum. The notation for the inverse temperature $\beta  = {\left( {kT} \right)^{ - 1}}$ is introduced. The excitation energies ${\varepsilon _n}$ of each state are equal to ${\varepsilon _n} = {I_H}\left( {1 - {n^{ - 2}}} \right)$. The statistical weight of an individual state in the sum in the formula \ref{Eq_Saha_sum} is modified according to the expression ${\tilde g_n} = {g_n}/\left( {{g_p}{g_e}} \right)$, where ${g_n} = 2{n^2}$, and ${\tilde g_n} = {n^2}$. The main feature in the new Equation~\ref{Eq_Saha_sum} are weight cut-off factors ${w_n}$, which describe the possibility of existence of hydrogen states with high quantum numbers $n$ in a high-temperature dense plasma.

Choosing a system of weight functions ${w_n}$ is one of the most difficult problems in modeling of dense plasmas. One advanced solution was implemented in the MHD EOS \citep{Mihalas_1988}. In this paper we estimate the partition function as it was proposed in the SAHA-S EOS framework (\citeauthor{Gryaznov_2004} \citeyear{Gryaznov_2004}, \citeyear{Gryaznov_2006}).

Before moving on to the properties and calculations of cut-off factors and the sum over excited states, let us point out the reason why we cannot do without them. Assuming ${w_n} = 1$, for all $n$, several problems arise. On the one hand, summing over infinite number of states in Equation~\ref{Eq_Saha_sum} results in a divergent sum at all temperatures, including low temperatures. That is, such a model cannot be used even in the trivial case of neutral hydrogen. On the other hand, if we restrict the sum to any finite number of states, then, as the temperature and density increases toward the center of the Sun, we obtain a prediction of a significant number of neutral hydrogen atoms in the center.

To obtain an alternative set of weight cut-off factors, we use the result obtained in different thermodynamic approach, the framework of the physical picture. In this case, we obtain an expression similar to Equation~\ref{Eq_Saha_sum}, and equivalent to the law of acting masses \citep{Ebeling_Kraeft_Kremp_1976}. According to this law, the concentration ratio is equal to a function of temperature:
\begin{equation}
\frac{{{n_H}}}{{n_e^{}n_p^{}}} = \mathchar'26\mkern-10mu\lambda _e^3 \cdot U.
\label{Eq_Saha2}
\end{equation}

\noindent
The function $U(T)$ will be called partition function (hereafter PF):
\begin{equation}
U = {e^{\beta {I_H}}}\sum\limits_{n = 1}^\infty  {{{\tilde g}_n}} {e^{ - \beta {\varepsilon _n}}}{w_n}.
\end{equation}

\noindent
In this expression, we denote the sum over quantum states of hydrogen as $Q$:
\begin{equation}
Q = \sum\limits_{n = 1}^\infty  {{{\tilde g}_n}} {e^{ - \beta {\varepsilon _n}}}{w_n}.
\label{Eq_Q}
\end{equation}

\noindent
The sum $Q$ is introduced in analogy with the Saha equation (\ref{Eq_Saha_sum}) and can be interpreted as the effective statistical weight of hydrogen taking into account excited states. 

Theoretical solutions have been obtained for the PF in the case of the hydrogen atom, two of which we will use in analyzing the equations of state. One of them is the Planck-Larkin (PL) and the other the Starostin-Roerich (SR) partition functions. They are defined in the next section.

Note that the analogy between the PF and the right part of the Saha equation (\ref{Eq_Saha_sum}) is superficial and based on external similarity of the obtained expressions. The PF is a result of elaborate quantum-statistical calculations. It is an integral over all possible states of the proton-electron pair, including not only classical hydrogen states, but also so-called scattered states. Expressions in the form of a sum over hydrogen states are only one possible form for PF. But in our case, we take advantage of this analogy and use the expressions of the weight factors ${w_n}$ from this function to obtain the corresponding expressions in the Saha equation.

It may seem that the sum (\ref{Eq_Q}) allows us to determine the populations of excited states. \cite{Potekhin_1996} discussed the difference between thermodynamic populations and those observed in optical experiments. In our case, following \cite{Rogers_1986}, in the low-density limit, using PL partition function, formal thermodynamic populations are given by the formula

\begin{equation}
\frac{{n_H^{(n)}}}{{{n_H}}} = \frac{{{{\tilde g}_n}{e^{ - \beta {\varepsilon _n}}}{w_n}}}{{\sum\limits_{n = 1}^\infty  {{{\tilde g}_n}} {e^{ - \beta {\varepsilon _n}}}{w_n}}}.
\label{Eq_Wn_PL}
\end{equation}

\noindent
\cite{Rogers_1986} also notes that the populations obtained in this way may differ markedly from the observated (optical) ones, which probably does not allow direct comparison with the populations estimated by \cite{Pradhan_2024}.
	
We should prevent from interpretation of the individual terms in the sum of Equation~\ref{Eq_Q} as the populations of a separate excited state. Each term in Equation~\ref{Eq_Q} is the sum of the contributions of bounded and free states (for more details, see \citealp{Rogers_1986}).  Only whole sum has the physical meaning  and the corresponding cut-off factors provide its convergence.

%%%%%%%%%%%%%%%%%%%%%%%%%%%%%%%%%%%%%%%%%
\section{Properties of Cut-off Factors and Partition Function}

The purpose of using cut-off factors is to obtain a finite sum in  Equation~\ref{Eq_Q} over an infinite number of excited states of hydrogen. It is important that the PF be a differentiable function of temperature, i.e., free of jumps and discontinuities.

The calculations are based on the key function ${w_n}$. This function is only a function of temperature, i.e., it does not depend on the density and concentration of particles. To calculate ${w_n}$, one also needs the ionization potentials of excited states ${I_n}$, which are equal to ${I_n} = {I_H}/{n^2}$ for hydrogen. But ultimately the cut-off factors $w\left( \alpha  \right)$ are a universal function of a single argument $\alpha  = {I_n}/(kT)$.

The Planck-Larkin partition function discussed in the introduction (commonly, and hereafter named PL) is widely used in the physical literature, for example \citealp{Ebeling_Kraeft_Kremp_1976, Krasnikov_1968, Kraeft_Kremp_Ebeling_Ropke_1986, Rogers_1996}, etc. The definition for the PL cut-off is
\begin{equation}
w_n^{\mathrm{PL}}\left( \alpha  \right) = 1 - {e^{ - \beta {I_n}}} - \beta {I_n}{e^{ - \beta {I_n}}}.	
\end{equation}

\noindent
According to \cite{Ebeling_Fortov_Filinov_2017} and \cite{Starostin_2009}, the PL function partially includes scattered states besides bound states of an atom. 
Another form of a cut-off factor is introduced in \cite{Starostin_2003,   Starostin_2006} which we name SR (after Starostin and Roerich). The SR cutoff factor ${w_n}$ is more complex than that of PL. It does not include scattered states. The SR partition function is used in SAHA-S equation of state \citep{Gryaznov_2006}: 
\begin{eqnarray} 
w_n^{\mathrm{SR}}\left( \alpha  \right)  =  1 & - & {e^{ - \beta {I_n}}}\left[ {4 - \frac{6}{{\sqrt \pi  }}{{\left( {\beta {I_n}} \right)}^{1/2}} + \frac{4}{{\sqrt \pi  }}{{\left( {\beta {I_n}} \right)}^{3/2}}} \right] + \nonumber \\
 & + &\frac{{\Gamma \left( {\frac{1}{2},\beta {I_n}} \right)}}{{\sqrt \pi  }}\left[ {3 - 4\beta {I_n} + 4{{\left( {\beta {I_n}} \right)}^2}} \right].
 \label{Eq_Wn_SR}
\end{eqnarray}

\noindent
Here the incomplete gamma function is	
\begin{equation}
\Gamma \left( {\frac{1}{2},z} \right) = \int\limits_z^\infty  {\frac{{{e^{ - t}}}}{{\sqrt t }}\mathrm{d}t.} 
\end{equation}

\noindent 
Formally, the given definitions are sufficient for solving the hydrogen ionization problem. In the following, we will consider the behavior of functions and discuss the difference in the results obtained.

%%%
\begin{figure} 
\centerline{\includegraphics[width=1.0\textwidth,clip=]{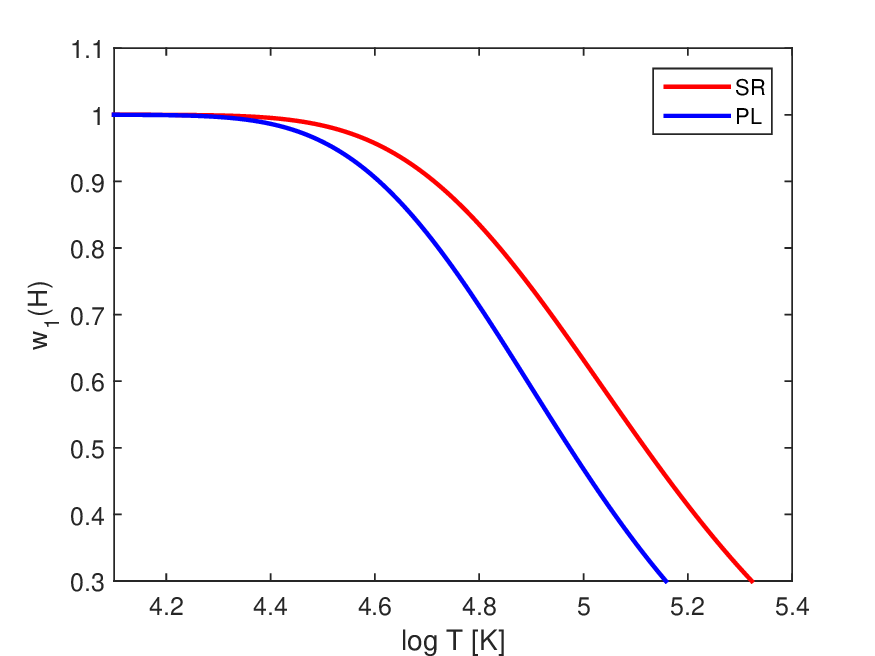}}
\caption{Cutoff factor $w_1$ for the ground state ($n=1$) of hydrogen in the PL and SR approximation. }
\label{Fig_w1}
\end{figure}
%%%

The general behavior of the cut-off factors is shown in Figure~\ref{Fig_w1}. Here, the factors are presented as functions of temperature, assuming that the ionization potential is equal to ${I_H}$, i.e., for the ground state of hydrogen. This figure is a universal description of cut-off factors. That is, to obtain multipliers for states with the number $n$, we only have to shift the value of the argument $\alpha $ of the function $w\left( \alpha  \right)$, assuming that the ionization potential of the excited state is smaller. An example of calculation of the corresponding ${w_n}$ at one temperature is given in the Appendix~\ref{Appendix_Wn}, Figure~\ref{Fig_Wn}.

The general behavior of the cut-off factors is quite simple. At low temperatures (high values of $\alpha $), $w = 1$. That is, at the temperature $T = 1.5\cdot 10^4$ K the factor for the ground state is 1, and makes no contribution to the ionization equation. As the temperature increases, $w$ starts to decrease, and the rate of decrease is asymptotically the same for both functions and is large enough to ensure convergence of the sum. The difference in the two models is that the magnitude of the ratio ${w^{\mathrm{SR}}}/{w^{\mathrm{PL}}} \to 4$ in the limit of high temperatures \citep{Starostin_2004}. Accordingly, the transition to the region of decrease in the case of the SR model occurs later than in the case of PL. As a result, the ${w^{\mathrm{SR}}} > {w^{\mathrm{PL}}}$ and SR multiplier for any state turns out to be always larger than for PL (Figure~\ref{Fig_w1}). On the other hand, the ratio of functions ${w^{\mathrm{SR}}}/{w^{\mathrm{PL}}}$ varies from unity at low temperatures, to 4 at high temperatures \citep{Gryaznov_2004}. This means that at the low temperatures one should not expect a difference in calculations with different models.

Let us return to the question about the factors $w$ for the excited states. States with high $n$ and small ionization potential $I_n={I_H}/{n^2}$ at any temperature will fall into the region of the decreasing tail of the cut-off factors, i.e., will cease to contribute to the sum over states. The only difference is in the value ${n^*}$, after which the excited states become negligibly small.
At low temperatures, states with $n < n*$ (excluding the ground state) appear uninhabited due to the Boltzmann factor ${e^{ - \beta {\varepsilon _n}}}$, and higher states with $n > {n^*}$ are cut off by $w$. As a result, the whole sum is determined only by the ground state $n = 1$. This effect can be called ``freezing" of excited states.

However, as the temperature increases,  ${w_n}$ of all states, including the ground state, decrease. The specificity of the cutoff factors model is that all states start to vanish very early, at low temperatures, but the dependence on $n$ is rather weak. As a result, the most important is the transition region from temperatures where $w=1$ to the region where all states, including the ground state, are cut off. For the SR function, this transition region is shifted toward high temperatures. As a result, taking into account excited states, the SR model predicts a noticeably higher weight of neutral hydrogen compared to the PL case. Moreover, this increase in the fraction of states is associated precisely with excited states. The contribution of individual states with small $n$ can be seen in Figure~\ref{Fig_Qn} in the Appendix~\ref{Appendix_Wn}.

Note that the choice of cut-off factors guarantees the exclusion of highly excited states, but it happens as delicately as possible. That is, the cut-off factor behaves with temperature in such a way as to ensure convergence of the sum. In reality, the exclusion of highly excited states can occur more quickly.

\begin{figure} 
\centerline{\includegraphics[width=1.0\textwidth,clip=]{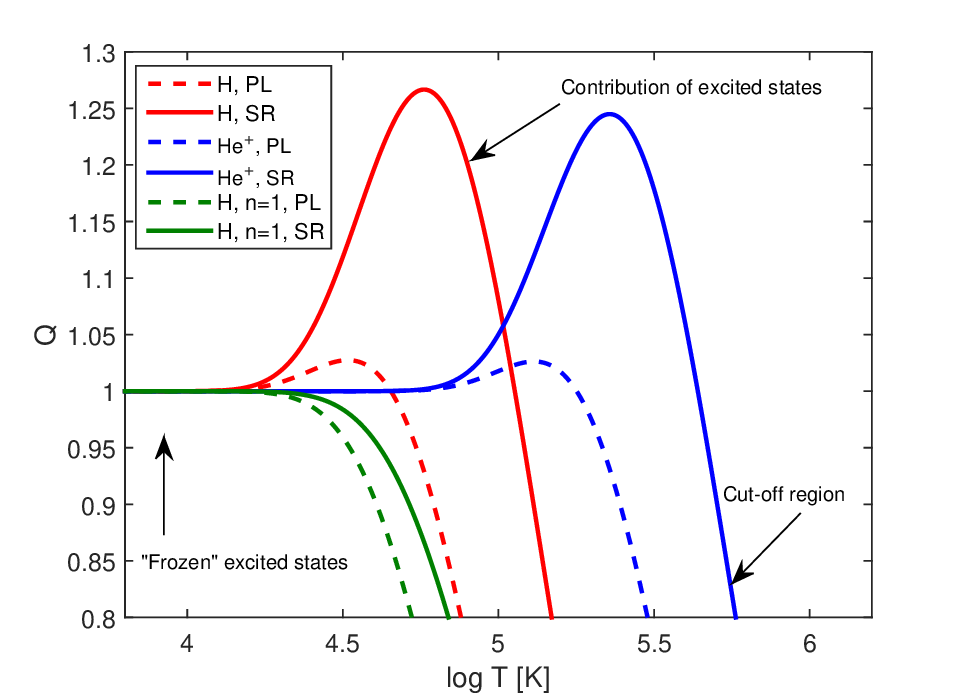}}
\caption{Partition function $Q$ calculated in SAHA-S EOS for hydrogen (\textit{red curves}) and helium (\textit{blue curves}). \textit{Green curves} show contribution of the main state ($n=1$) of hydrogen atom. \textit{Solid curves} are for SR approach, \textit{dashed} – for PL.  }
\label{Fig_Q}
\end{figure}

Let us proceed to the calculation of the sum $Q$ (Equation~\ref{Eq_Q}) by states using the SR and PL functions. An example of such a calculation is shown in Figure~\ref{Fig_Q}. 

%%%
%%%

The terms in the sum (\ref{Eq_Q}) are the product of the functions ${w_n}$ and the Boltzmann multiplier ${e^{ - \beta {\varepsilon _n}}}$. The behavior of the sum $Q$ depends on where the transition to the cutoff region occurs. It is this region and ionization in it that will be the subject of our consideration. 

For each state, ${w_n}$ only decreases with increasing temperature, but the total sum and PF of hydrogen can grow, reaching a maximum in the intermediate region. This effect is observed even with PL, albeit to a lesser extent, and even more so with SR (see Figure~\ref{Fig_Q}). 

Hydrogen ionization starts at low temperatures and high values of $\beta {I_H} \sim 10$. Under these conditions, the difference between $w_{PL}$ and $w_{SR}$ is still minor, since both functions are equal to unity. However, as the temperature increases, the functions show a significant difference between the descriptions within the different models. 

The maximum of $Q_H^{\mathrm{SR}}$ is 1.25 compared to the maximum of $Q_H^{\mathrm{PL}}$ around 1.05. The maximum of  $Q_H^{\mathrm{SR}}$ is noticeably shifted towards higher temperatures. The SR-function is always larger than the PL-function $Q_H^{\mathrm{SR}} > Q_H^{\mathrm{PL}}$. Note that the cut-off functions PL or SR have been originally proposed for hydrogen. However they can also be applied to hydrogen-like ions. On the figure \ref{Fig_Q}, the $Q$ -profile for $He^{+}$ are shown. In sense of $Q\left( \alpha  \right)$, the partition function should be the same for all hydrogen-like ions.

In the general case, the cut-off factors do not guarantee absence of inverse recombination of hydrogen because they depend on temperature only. Saha equation with these factors predicts hydrogen becomes neutral while density increases at any fixed temperature. 
	
But we checked the ionization of hydrogen at the center of the Sun. 
PFs predict the disappearance with temperature of all states, including the ground state. As the temperature increases in the central regions of the Sun, $w$ predicts rather small weights for the ground and other hydrogen states, since the argument $\alpha $ there is equal to 0.01.  As a result, the fraction of neutral hydrogen below the convection zone is small. Appendix~\ref{Apendix_H_ionization_center} shows fraction of hydrogen atoms in the central part of the Sun for the both cases PL and SR.

On the other hand, density effect and limitation of number of levels could affect the calculation of partition function. In Appendix~\ref{Appendix_Mott}, we demonstrate  the partition function with finite number of levels by applying the Mott condition (see, for example, \citealp{Ebeling_2012, Ebeling_2020}). The effect is insignificant compared to the difference between the PL and SR functions under solar conditions.

The magnitude of the sum $Q$ can be interpreted as the effective statistical weight in the Saha equation (\ref{Eq_Saha2}). Thus, having the calculation of the sum $Q$  in both models, we can proceed to the analysis of hydrogen ionization calculated within the SAHA-S equation.

%%%%%%%%%%%%%%%%%%%%%%%%%%%%%%%%%%%%%%%%%
\section{Hydrogen Ionization in Solar Conditions: Results of Calculation}

\begin{figure} 
\centerline{\includegraphics[width=1.0\textwidth,clip=]{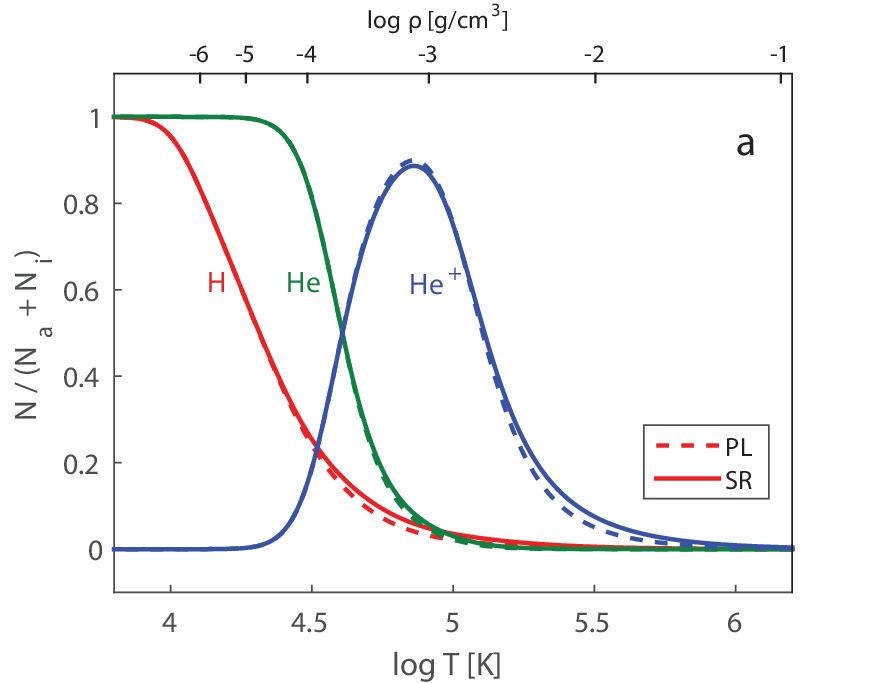}}
\centerline{\includegraphics[width=1.0\textwidth,clip=]{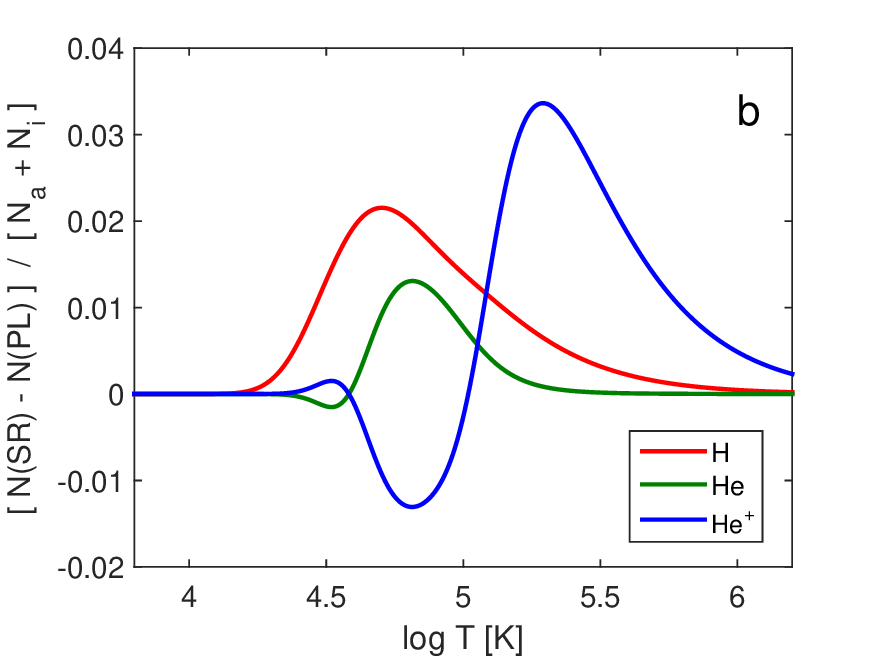}}
\caption{Hydrogen and helium ionization at the $\left( {T,\rho } \right)$ points of solar model. \textit{Red curves} are for hydrogen atom, \textit{green} – for helium atom, and \textit{blue} – for ion $\mathrm{He}^+$. (a) \textit{Dashed curves} are obtained using PL partition function, \textit{solid} – using SR partition function. (b) Difference between ionization in SR and PL approaches.}
\label{Fig_ionization_HHe}
\end{figure}

Figure~\ref{Fig_ionization_HHe} demonstrates hydrogen and helium ionization calculated at the $\left( {T,\rho } \right)$ points of the solar model.  Ionization is shifted towards higher temperature in SR-case in comparison with PL-case. Hydrogen ionization in PL-case is systematically larger then SR-case and the difference (red solid line in Figure~\ref{Fig_ionization_HHe}(b)) has the same sign at all temperatures. There are no differences between PFs at low temperatures ($\log T < 4.2$); $Q$ sum equals 1 for both PFs. At higher temperature, the ionization difference is slowly going to zero because both models predict complete ionization here. The difference in hydrogen ionization is remarkable over helium ionization zone with $\log T \simeq 5$. It will cause significant effect on helium hump (see below).

In Figure~\ref{Fig_ionization_HHe}, the helium ionization is also shown. The helium ionization is considered both in PL and SR-model. Generally, the helium ionization differences are similar to those of hydrogen ionization, but fine details in Figure~\ref{Fig_ionization_HHe}(b) look specific for two-electrons ionization. 
We should note that there is no rigorous background for application of the PL and the SR expressions to the He atom. 
But in absence of convergent partition function for helium atom, in present SAHA-S calculations, the same form was used for all atomic components.

%%%%%%%%%%%%%%%%%%%%%%%%%%%%%%%%%%%%%%%%%%%%%%%%
\section{The First Adiabatic Exponent Profile}

\begin{figure} 
\centerline{\includegraphics[width=1.0\textwidth,clip=]{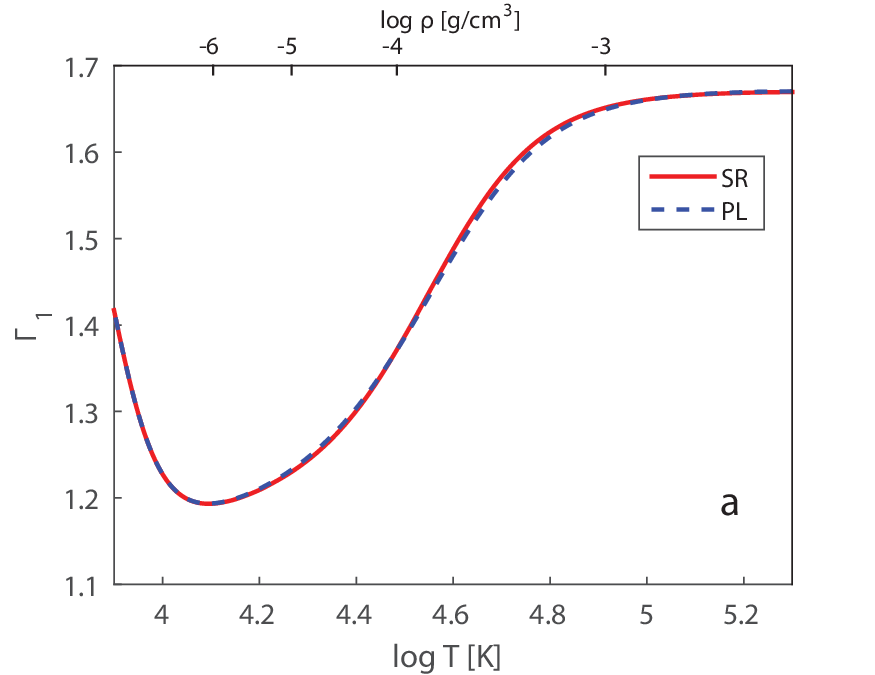}}
\centerline{\includegraphics[width=1.0\textwidth,clip=]{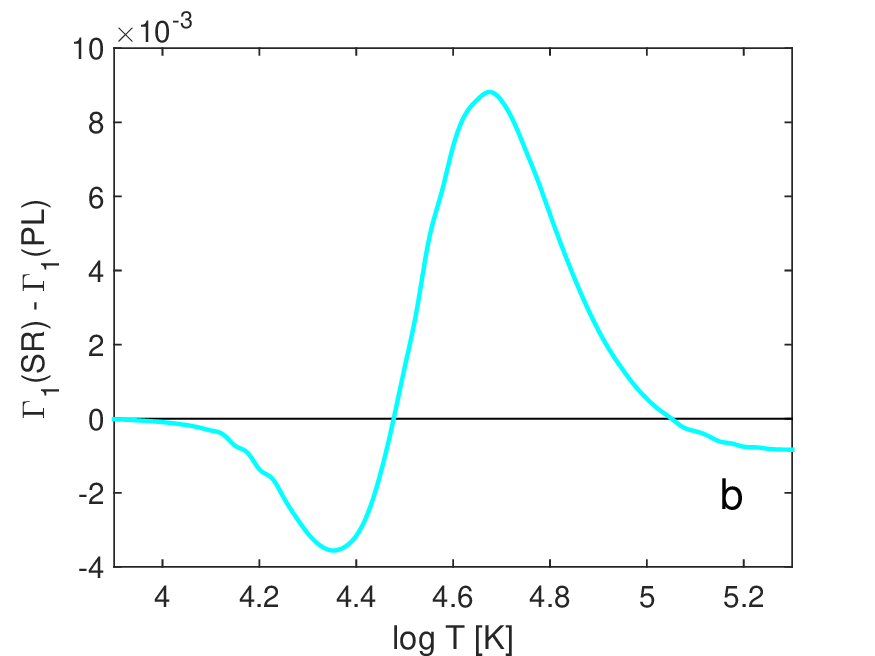}}
\caption{${\Gamma _1}$  profiles in SR and PL approaches for pure hydrogen plasma (a) and difference between them (b). ${\Gamma _1}$ are computed at points $(T,\rho)$ from solar model.}
\label{Fig_G1_H_SR_PL}
\end{figure}

Figure~\ref{Fig_G1_H_SR_PL} shows effect of the PF on ${\Gamma _1}$ profiles for pure hydrogen plasma. 
The profile in the PL case appears to be shifted toward higher temperature at the $\log T \simeq 4.8$.
Because of the ionization delay, one would expect the ${\Gamma _1} \left( \mathrm{SR} \right)$ profile to be shifted toward high temperatures. But this would be the case if ionization shifts uniformly over the entire region. In this case, the profile shift would lead to the fact that ${\Gamma _1}\left(\mathrm{PL} \right) < {\Gamma _1}\left( \mathrm{SR} \right)$ in the decline region, and after passing the minimum in ${\Gamma _1}$, the sign of the difference would be reversed. 

However, this is not the case in our comparison. Before reaching the minimum of ${\Gamma _1}$ there is no difference between the two cases. This is because hydrogen ionization begins at temperatures lower than that when the sum variation begins in $Q\left( \mathrm{PL} \right)$. That is, the description of the initial stage of ionization is the same in the PL and SR models. The differences between the models begin just in the region of the ${\Gamma _1}$ minimum or even at higher temperatures. Note also that the ionization of hydrogen at the ${\Gamma _1}$ minimum is only 20\%.

In the region where there is an increase of the hydrogen sum in the SR case, ionization is delayed. As a result, in the region after the ${\Gamma _1}$ minimum, we have the situation ${\Gamma _1}\left( \mathrm{PL} \right) > {\Gamma _1}\left( \mathrm{SR} \right)$, which is opposite to what is expected above. In Figure~\ref{Fig_G1_H_SR_PL}(b) (bottom panel), this region corresponds to negative values of the difference ${\Gamma _1}\left( \mathrm{SR} \right) - {\Gamma _1}\left( \mathrm{PL} \right)$. 

With further temperature increase, the pattern is reversed, and again ${\Gamma _1}\left( \mathrm{PL} \right) < {\Gamma _1}\left( \mathrm{SR} \right)$. This corresponds to the theory of isoentropic curves in the ionization regions. Let us choose points A and B for which hydrogen is neutral and fully ionized, respectively. At these points, the pressure difference between the SR and PL models is zero (see Figure~\ref{Fig_dP_SR_PL} in Appendix~\ref{Appendix_dP}). Then we can write
\begin{equation}
\int\limits_A^B {\delta {\Gamma _1}\mathrm{d}\log \rho }  = \int\limits_A^B {\mathrm{d}\left( {\delta \log P} \right)}  = \left. {\delta \log P} \right|_A^B = 0.
\end{equation}

\noindent
The integral is taken along the adiabat and can be taken also by the temperature:
\begin{equation}
\int\limits_A^B {\delta {\Gamma _1}{{\left. {\frac{{\partial \log \rho }}{{\partial \log T}}} \right|}_S}\mathrm{d}\log T}  = \int\limits_A^B {\delta {\Gamma _1}{{\left( {{\Gamma _3} - 1} \right)}^{ - 1}}\mathrm{d}\log T}  = 0.
\end{equation}

\noindent
Thus, a smaller ${\Gamma _1}$ in one region must be compensated by a region of the opposite sign. 
As a result, we observe a picture not just of ${\Gamma _1}$ displacement, but more complex deformation of the profile in the region of ${\Gamma _1}$ growth, in the form of winding model profiles. Amplitude of the difference between SR and PL reaches ${10^{ - 2}}$. 
The difference between $\Gamma_1$ in SR and PL approaches does not sensitive to points $(T,\rho)$ of a solar model.  We examine this effect for several solar models in Appendix~\ref{Appendix_models}.

In Figure~\ref{Fig_G1_H_SR_PL}(b), we see the deviation of the ${\Gamma _1}$-difference from zero at temperatures $\log T > 5$. The reason for this, in our opinion, is the Coulomb effect. But the consideration of this effect is beyond the scope of this paper.

\begin{figure} 
\centerline{\includegraphics[width=1.0\textwidth,clip=]{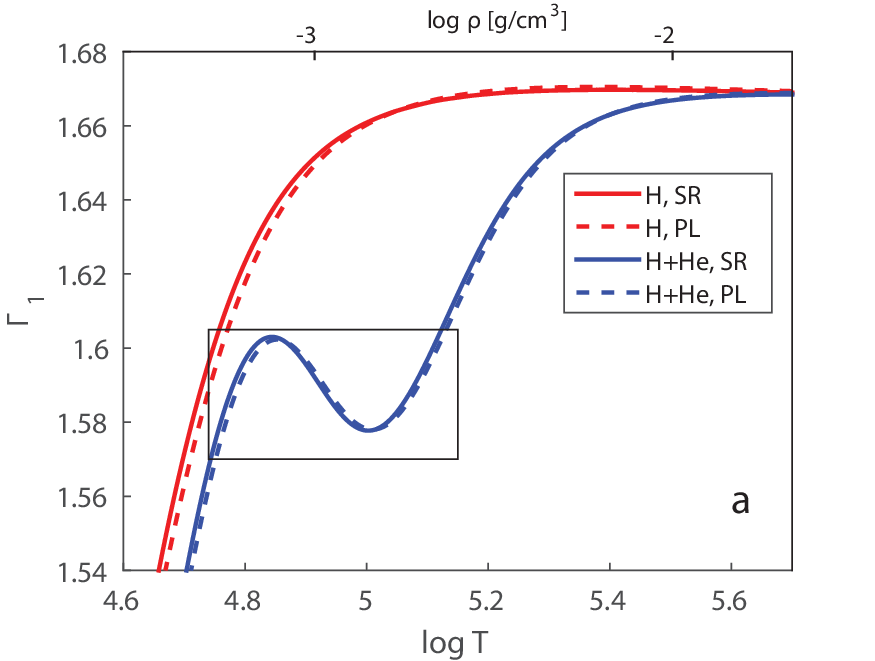}}
\centerline{\includegraphics[width=1.0\textwidth,clip=]{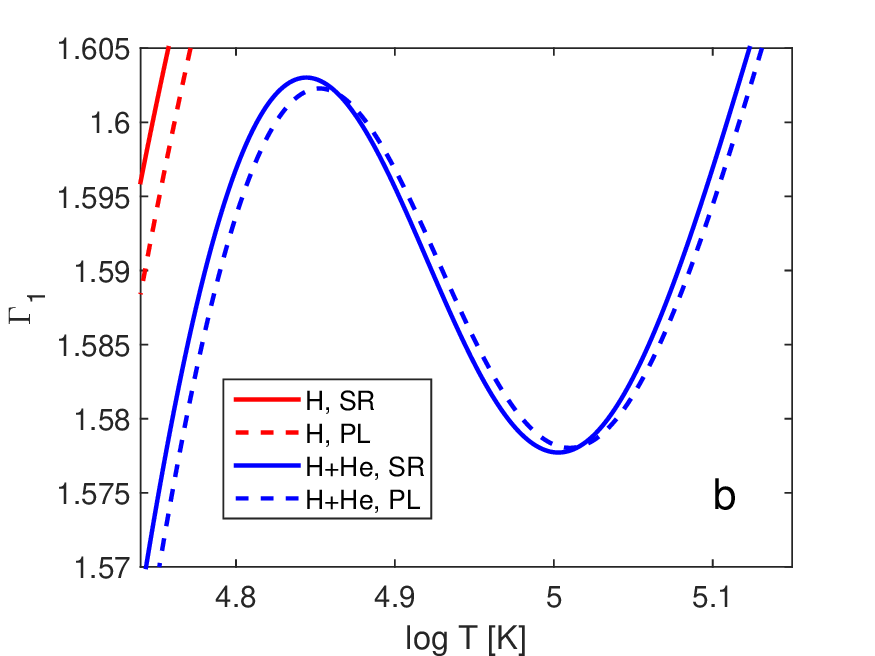}}
\caption{${\Gamma _1}$  profiles in H-He plasma ($X=0.75$, $Y=0.25$) in SR and PL approaches at different scales (a) and (b).}
\label{Fig_G1_HHe_SR_PL}
\end{figure}

The same effect of ${\Gamma _1}$-disturbance appears for all H-like ionizations. Figure~\ref{Fig_G1_HHe_SR_PL} demonstrates the effect of hydrogen shifting on the position of the helium hump. With respect to the ${\Gamma _1}$ profile, the helium hump shifts almost isomorphically toward high temperatures in the transition from the SR to the PL model. This direction of the shift is not consistent with the shift of the ionization zones, including the helium ionization shown in Figure~\ref{Fig_ionization_HHe}. Helium ionization, like hydrogen, occurs at higher temperatures in the SR model. However, the shifted ${\Gamma _1}$ profile for a purely hydrogen plasma shows exactly the same shift in temperature as in the case of helium ${\Gamma _1}$ profiles.

The explanation of this effect is related to the interaction of hydrogen and helium ionizations through the number of free electrons. In the case of SR model, hydrogen ionization occurs at higher temperatures. That is, the number of free electrons is smaller than in the case of PL model. This leads to the fact that the ionization of helium is easier because the fewer free electrons, the easier it is for the corresponding ion to ionize. As a result, in the SR model, the perturbation of ${\Gamma _1}$ due to helium is shifted toward lower temperatures compared to the case of the PL model.

%%%%%%%%%%%%%%%%%%%%%%%%%%%%%%%%%%%%%%%%%%%%%%%%%%%%%%%%%%%%%%%%%%
\section{Conclusion}

Two models of cut-off factors describing the contribution of excited states have been used to analyze a pure hydrogen plasma under solar conditions. The SR model predicts a larger contribution of bound states of hydrogen than the PL model. The main result is that in the case of the SR model, hydrogen ionization occurs deeper (at higher temperatures) than in the case of the PL-function. The proton and free electron number curves are shifted toward higher temperatures for the SR case. 
The shift of the ionization region is fully reflected in the pressure difference in the two models. The pressure in the SR case is always lower than in the PL case, and the SR-PL difference represents a bell-shaped depression (see Appendix~\ref{Appendix_dP}). 

However, from the systematic temperature shift of ionization, two corollaries emerge. 

The first one concerns the behavior of ${\Gamma _1}$ profiles. The perturbation of ${\Gamma _1}$  under the considered conditions does not lead to a simple shift of its profile in depth. In fact, the ${\Gamma _1}$ profiles in the two models wind up on each other. While in the region of the ${\Gamma _1}$ minimum the transition to the SR-function prolongs smaller ${\Gamma _1}$ to the region of high temperatures (ionization is delayed), the opposite effect occurs with further temperature increase. The ${\Gamma _1}$ in the SR-model reaches the limit values faster for a fully ionized plasma. This effect follows from the theory of adiabatic profiles, and is clearly shown on the example of pure hydrogen.  As a result, the total ionization profile of ${\Gamma _1}$ does not shift much with the degree of ionization, but rather deforms in its (hotter) wing, becoming steeper. This effect is considered on the example of a purely hydrogen plasma. 

Another consequence concerns the interaction of hydrogen and helium ionizations at solar conditions. Helium ionization occurs against the background of the ongoing incomplete ionization of hydrogen. It was found that if hydrogen ionization is delayed, helium ionization is enhanced on the contrary and occurs at lower temperatures. Conversely, easier ionization of hydrogen leads to more difficult ionization of helium. 

An accurate description of the perturbation of the ${\Gamma _1}$ profile due to the contribution of excited states for the ionization of ions other than hydrogen and helium requires further study. This is important for the method of decomposition by $Z$ contributions \citep{Baturin_2022}.
 
The model for cut-off factors, borrowed from the PF theory in the physical picture, is able to provide a framework for describing the ionization of hydrogen and helium at solar conditions.  In the outer layers of the Sun, where hydrogen ionization process is dominant, we need enhancement of models of the hydrogen atom to improve the EOS description. The transition from one model of hydrogen ionization to another is able not only to strongly perturb the ${\Gamma _1}$ profile in the outer layers, but also indirectly affects the position of the helium ionization zone. Both features play a fundamental role in the inversion procedures, as well as in the construction of the seismic model of the Sun.

%%%%%%%%%%%%%%%%%%%%%%%%%%%%%%%%%%%%%%%%%%%%%%%%%%%%%%%%%%%%%%%%%%%%%%%%%%%
%% Appendix
%%%%%%%%%%%%%%%%%%%%%%%%%%%%%%%%%%%%%%%%%%%%%%%%%%%%%%%%%%%%%%%%%%

\appendix

%%%%%%%%%%%%%%%%%%%%%%%%%%%%%%%%%%%%%%%%%%%
\section{Hydrogen Ionization in Other Equations of State}
\label{Appendix_EOS}

Ionization of hydrogen is different generally in various equations of state. Comparison of SAHA-S EOS with selected equations of state was performed previously by \cite{Gryaznov_2016} along isohores. In contrast, we show comparison at the points of solar model. We examine equations of state widely used in astrophysics, i.e. OPAL \citep{Rogers_Nayfonov_2002} and FreeEOS \citep{Irwin_2012}.
	
\begin{figure} 
\centerline{\includegraphics[width=1.0\textwidth,clip=]{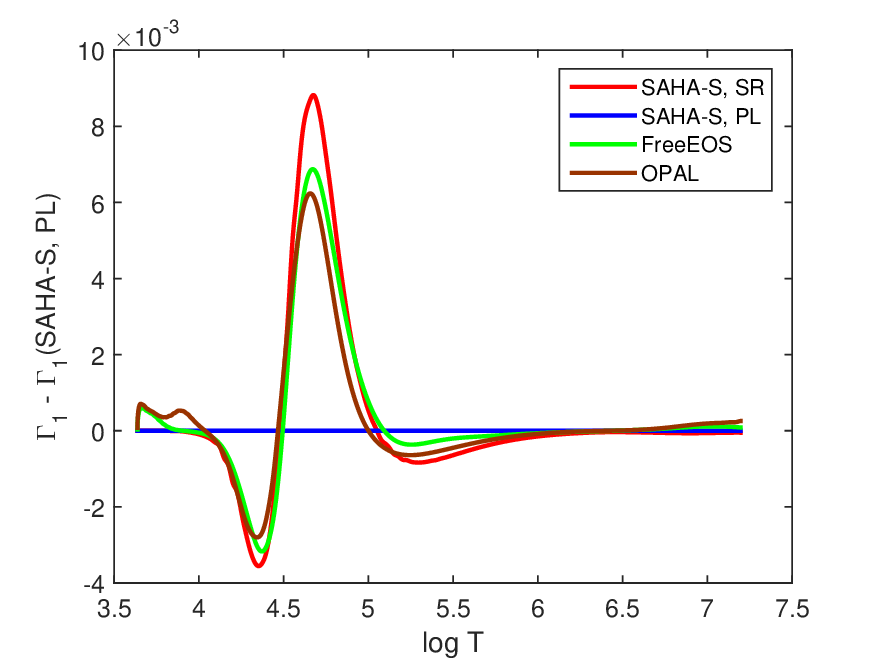}}
\caption{Deviations of $\Gamma_1$ for hydrogen plasma computed in various EOS from $\Gamma_1$ in SAHA-S with PL partition function.}
\label{Fig_dG1_eos}
\end{figure}

Figure~\ref{Fig_dG1_eos}  shows  differences of $\Gamma_1$  for hydrogen plasma. FreeEOS and OPAL curves lie between SR and PL ones, closer to SR, in the region of hydrogen ionization ($\log T\sim 4-5$). Their maximal deviation from PL curve is about $6\cdot 10^{-3}$. The deviations in other regions are smaller $10^{-3}$.
	
%%%%%%%%%%%%%%%%%%%%%%%%%%%%%%%%%%%%%%%%%%%%%%%%%%%%%%%%%%%%%%%%%%
\section{Cut-off Factors and Partition Functions for Several States} 
\label{Appendix_Wn}  

\begin{figure} 
\centerline{\includegraphics[width=1.0\textwidth,clip=]{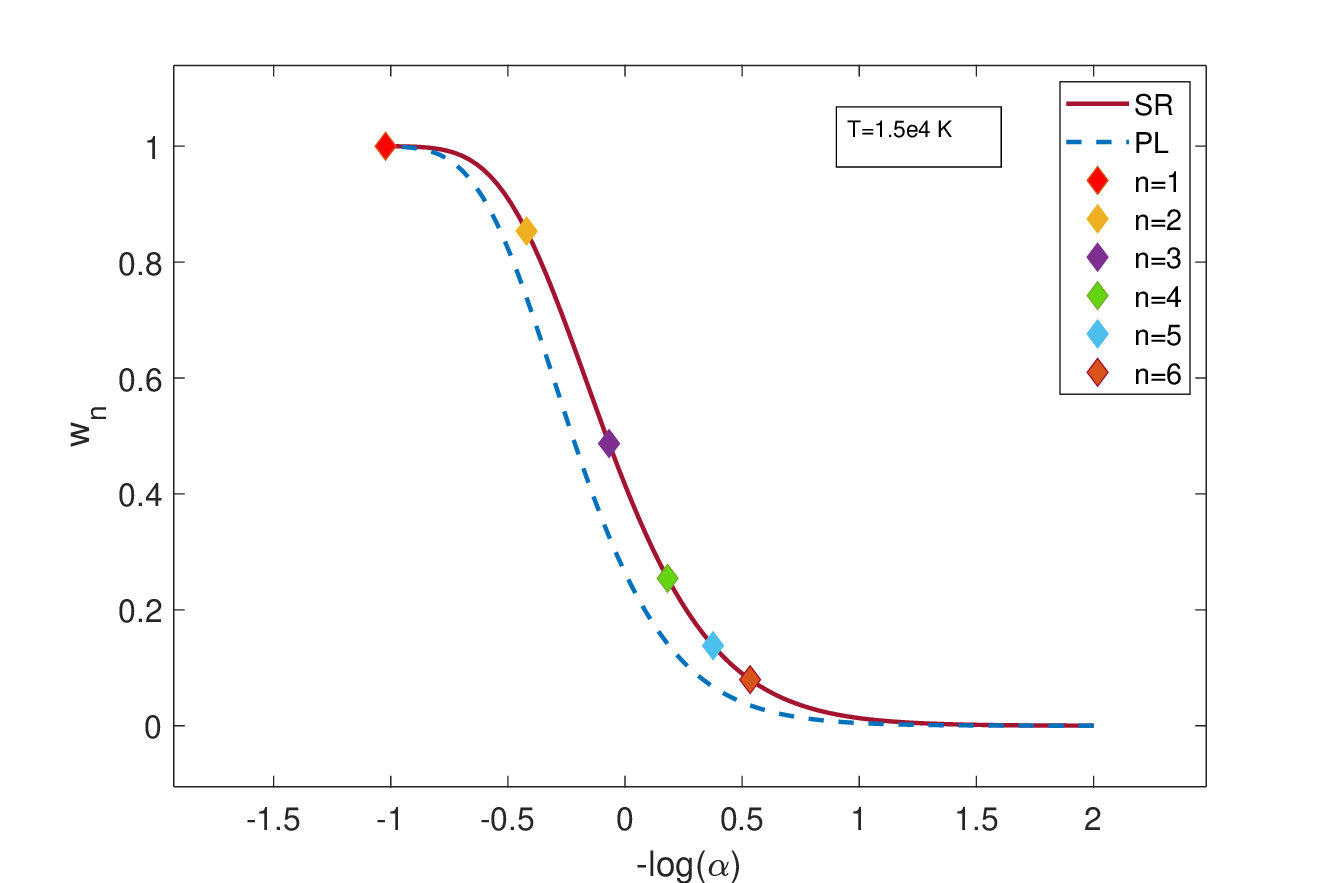}}
\caption{Cutoff factors ${w_n}$ according expressions \ref{Eq_Wn_PL} and \ref{Eq_Wn_SR}. An argument $\alpha=I_n/kT$, where $I_n=I_{\mathrm H}/n^2$ is ionization energy for state n. The curves on this plot are independent of the temperature, but the values $w_n$ depend and are shown for $T=10^4$~K. }
\label{Fig_Wn}
\end{figure}

Figure~\ref{Fig_Wn} shows the values of the weights of excited states for a fixed temperature. A sufficiently low temperature $T = 1.5 \cdot {10^4}K$ was chosen as an example. The weight factor for the ground state of hydrogen turns out to be close to unity. The values of other excited states were calculated for the same temperature using Expression (\ref{Eq_Wn_SR}). The ionization potentials for excited states are significantly lower than those of the ground state. Therefore, the calculation results turn out to be located at lower values of  ${\alpha _n}$, i.e. shifted to the right on the graph. The values of the weight factors of excited states quickly fall into the “cut-off” section of the weight function, even if the ground state continues to be equal to unity.

\begin{figure} 
	\centerline{\includegraphics[width=1.0\textwidth,clip=]{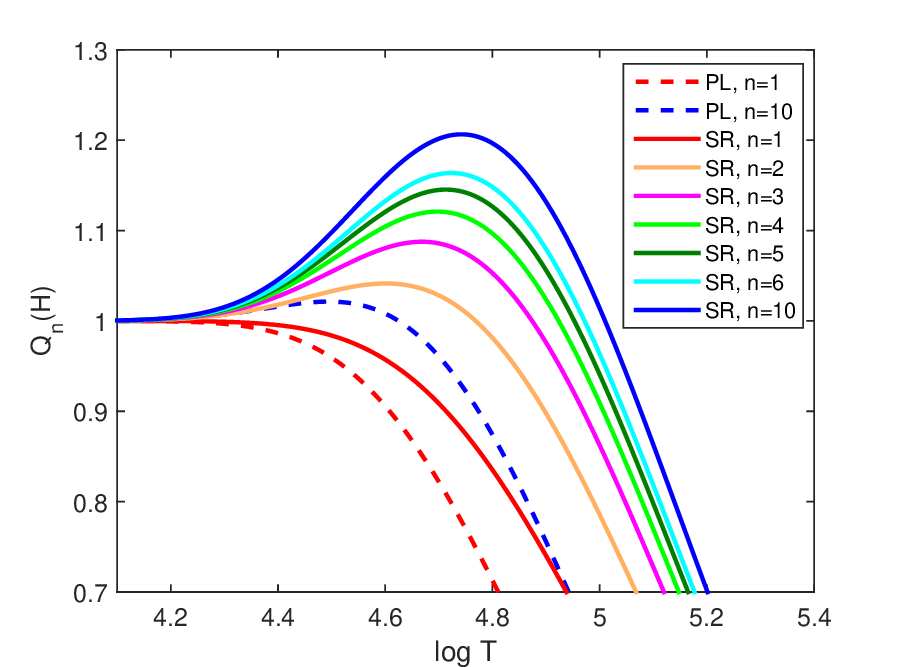}}
	\caption{Partition function $Q$ calculated for several limit numbers $n$. }
	\label{Fig_Qn}
\end{figure}

Figure~\ref{Fig_Qn}  shows the contribution of each subsequent excited state to the sum $Q$  compared to the sum calculated with a smaller number of states. For the SR model, the contribution of excited states is quite significant compared to the sum over the ground state, reaching 30\% in the region of the maximum sum $\log T = 4.8$. Excited states in the case of PL make a noticeable contribution, but it does not exceed 10\% of the ground state, and the sum itself does not exceed 1.04.

%%%%%%%%%%%%%%%%%%%

\section{Fraction of Hydrogen Atoms in the Central Part of the Sun}
\label{Apendix_H_ionization_center}

\begin{figure} 
	\centerline{\includegraphics[width=1.0\textwidth,clip=]{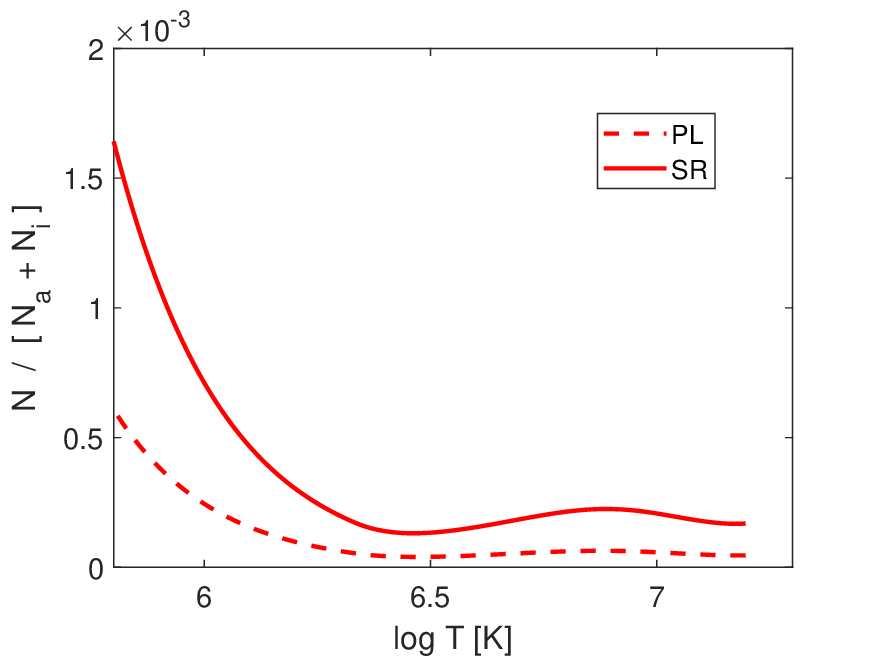}}
	\caption{Fraction of hydrogen atoms in the central part of the Sun.}
	\label{Fig_H_ioniz_center}
\end{figure}

The fraction of neutral hydrogen below the convection zone is plotted in Figure~\ref{Fig_H_ioniz_center} for the both cases PL and SR.  A part of neutral hydrogen is small as $10^{-4}$.

%%%%%%%%%%%%%%%%%%%%%%%%%%%%%%%%%%%%%%%%%%%%%%%%%%%%%%%%%%%%%%%%%%%%%%%
\section{Discussion on Mott Condition}
\label{Appendix_Mott}

The functions PL and SR depend only on temperature and are independent of density. In this section, we consider the question of how good this approximation is under the conditions of the convective zone of the Sun. At high densities, when the average distance between atoms is smaller than their sizes, the atoms are destroyed. This condition is called the Mott condition (see, e.g., \citealp{Ebeling_2012}). The radius of a hydrogen atom in an excited state with number $n$ can be estimated by the formula
	
\begin{equation}	
	a_n=n^2a_{\mathrm{B}} \, , 
\end{equation}	
	
\noindent where $a_{\mathrm{B}}$ is the Bohr radius. This equality allows us to estimate the critical density $\rho_n$, above which atoms with the level $n$ and higher do not exist:
	
\begin{equation}
	\rho_n=\frac{m_{\mathrm{H}}}{(4/3)\pi a_{n}^{3}} \, .	
	\end{equation}
	
\begin{figure} 
\centerline{\includegraphics[width=1.0\textwidth,clip=]{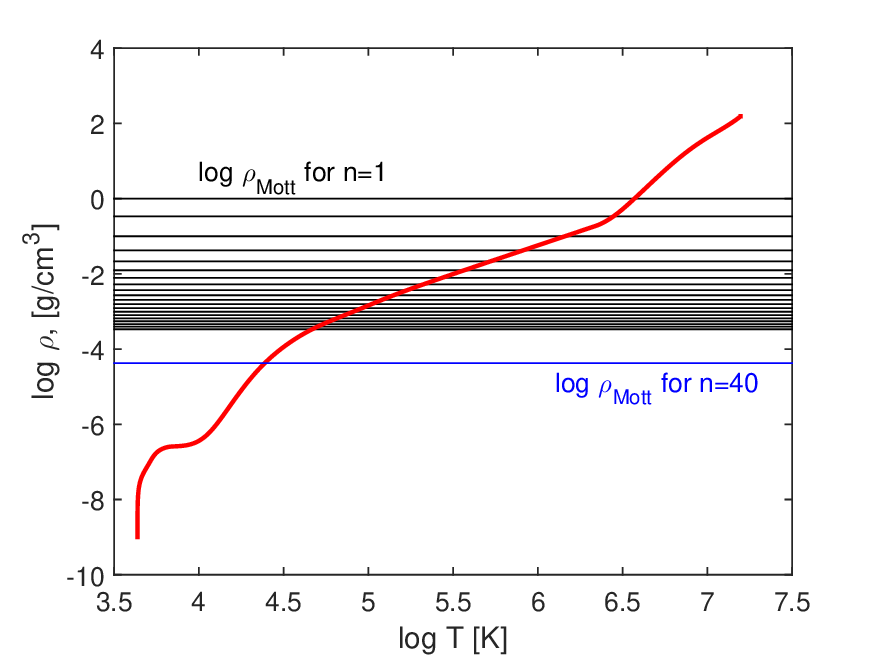}}
\caption{Densities corresponding to the Mott condition for $n=1-20$ (\textit{black horizontal lines}), $n=40$ (\textit{blue line}), as well as the temperature and density in the solar model  (\textit{red curve}).}
\label{Fig_T_rho_Mott}
\end{figure}

\noindent Here $m_{\mathrm{H}}$ is the mass of the hydrogen atom. In Figure~\ref{Fig_T_rho_Mott}, the critical densities for $n$ from 1 to 20 are shown by black horizontal lines. The difference between the densities $\rho_n$ also decreases with increasing $n$.	
	
This figure also shows the temperature and density in the solar model (red curve). According to the Mott condition, there are no neutral atoms at $\log  T>6.58$, they are destroyed due to the high density. At lower temperatures, $\log T=6.45-6.58$, only atoms in the ground state ($n=1$) can exist. At $\log T=6.16-6.45$, atoms exist in states $n=1$ and 2, and so on. 
	
\begin{figure} 
\centerline{\includegraphics[width=1.0\textwidth,clip=]{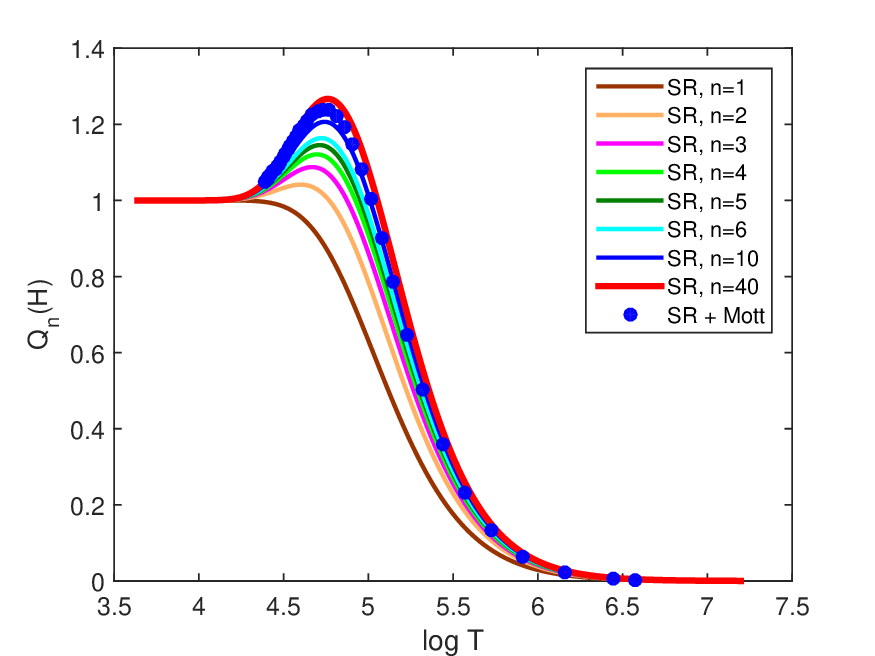}}
\caption{Partition function SR calculated for different numbers of levels (\textit{curves}), and also taking into account the Mott condition (\textit{blue circles}).}
\label{Fig_Qn_Mott}
\end{figure}

We calculate the partition function SR, leaving only the limited number of levels. The result is shown in Figure~\ref{Fig_Qn_Mott} by blue circles. Taking into account the bound states using Mott conditions does not significantly affect the value of the partition function Q(H). 
	
%%%%%%%%%%%%%%%%%%%%%%%%%%%%%%%%%%%%%%%%%%%%%%%%%%%%%%%%%%%%%%%%%%%
\section{Influence of Partition Function on Pressure}
\label{Appendix_dP}

\begin{figure} 
\centerline{\includegraphics[width=1.0\textwidth,clip=]{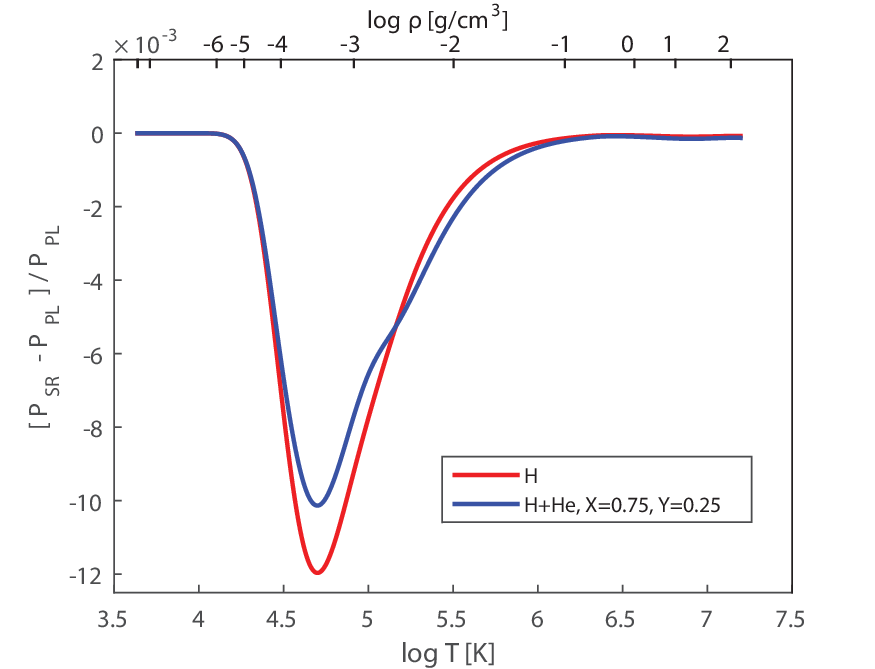}}
\caption{Difference between pressures computed in SR and PL approaches.}
\label{Fig_dP_SR_PL}
\end{figure}

Figure~\ref{Fig_dP_SR_PL} shows pressure differences between PL and SR cases for pure hydrogen and hydrogen-helium mixture. The pressure in the SR case is always lower than in the PL. The difference between the pressures in the ionization region of hydrogen and helium reaches one percent.

%%%%%%%%%%%%%%%%%%%%%%%%%%%%%%%%%%%%%%%%%%%%%%%%%%%%%%%
\section{Sensitivity to Solar Models}
\label{Appendix_models}

In this section, we examine if our result depends on points $(T,\rho)$. There are many standard and non-standard solar models in up-to-date literature (see e.g. review by  \citealp{Christensen-Dalsgaard_2021}). Our main result is obtained for  standard solar model computed with SAHA-S equation of state for high-Z solar abundances, i.e. mass fraction of elements heavier than helium is $Z=0.018$ \citep{Ayukov_Baturin_2017}. Now we consider, as an example, a solar model computed for low-Z abundances $Z=0.0136$ \citep{Ayukov_Baturin_2017}. We consider also Model~S by \cite{Christensen-Dalsgaard_1996}, a high-Z model, computed with OPAL equation of state, other nuclear reactions etc.

\begin{figure} 
	\centerline{\includegraphics[width=1.0\textwidth,clip=]{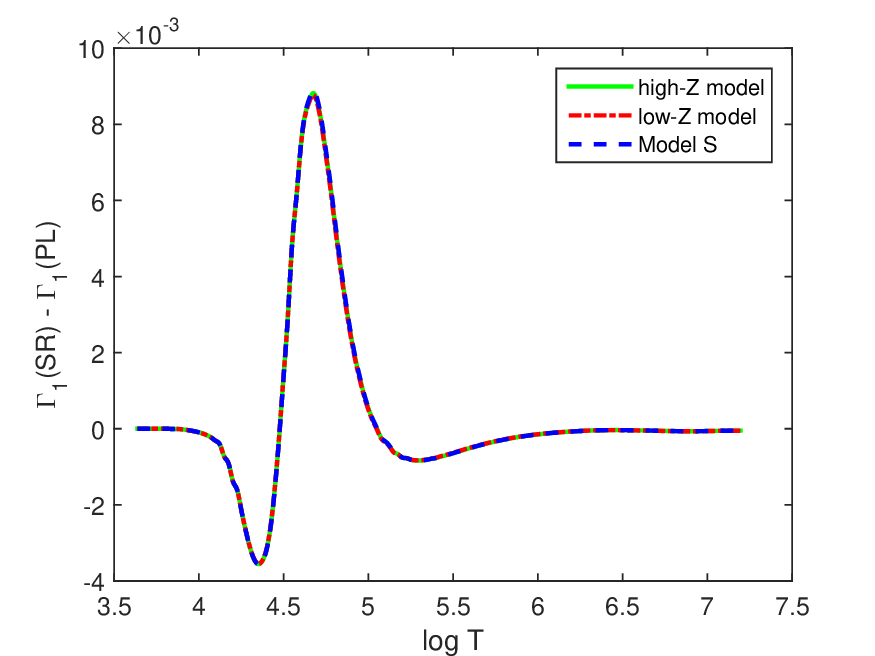}}
	\caption{Difference between $\Gamma_1$ for hydrogen plasma computed using SR and PL partition functions at points $(T,\rho)$ of various models.}
	\label{Fig_dG1_SR_PL_models}
\end{figure}

Figure~\ref{Fig_dG1_SR_PL_models} shows the difference between $\Gamma_1$ for hydrogen plasma computed using SR and PL partition functions at points $(T,\rho)$ of these three models. The curves are not distinguishable at the scale of the figure. Thus, the effect of partition function on adiabatic exponent is the same at points of various solar models.

%%%%%%%%%%%%%%%%%%%%%%%%%%%%%%%%%%%%%%%%%%%%%%%%%%%%%%%%%%%%%%%%%%%%%%%%%%%
%% Acknowledgements
%
\begin{acks}[Acknowledgments] All our research is based on the enormous contribution of our colleague, friend and teacher A.N. Starostin (1940-2020) to the quantum-statistical theory of the equation of state. \end{acks}

%% Available additional data environments:
%% required: authorcontribution, fundinginformation, dataavailability
%% optional: materialsavailability, codeavailability
\begin{authorcontribution}
V.A.B. proposed the idea of the thorough analysis of hydrogen ionization, from the atomic partition function to the ${\Gamma _1}$-behavior. The methodology of the paper, including theoretical, computational and editorial framework was performed by V.A.B.
A.S.V. was responsible for thermodynamics and solar modeling computations. 
A.V.O. and A.B.G. provide necessary numerical and theoretical computations of the partition functions. A.V.O. is mostly involved into writing and textual preparing of the paper.
V.K.G. and I.L.I. provided all necessary computations in the frames of SAHA-S EOS with PL and SR partition functions. They also provide a theoretical basis to the  Starostin-Roerich partition function.
W.D. provided the theoretical arguments for the combination of the chemical and physical pictures of the EOS.
All the co-authors contributed to the interpretation of the results and were involved in the discussions.

\end{authorcontribution}
\begin{fundinginformation}
The study by V.K. Gryaznov is conducted under the government contract for fundamental research registration number 124020600049-8.
\end{fundinginformation}
\begin{dataavailability}
No datasets were generated or analysed during the current study.
\end{dataavailability}
\begin{ethics}
\begin{conflict}
The authors declare no competing interests.
\end{conflict}
\end{ethics}

%%% %%%%%%%%%%%%%%%%%%%%%%%%%%%%%%%%%%%%%%%%%%%%%%%%%%%%%%%%%%%
%% Bibliography
%
% Using BibTeX
%
\bibliographystyle{spr-mp-sola}
\bibliography{bibliography}  
%
% Without BibTeX 
% \begin{thebibliography}{}
% \bibitem[\protect\citeauthoryear{Author}{Year}]{key}
%   <bibliographical entry>
%
% \bibitem[\protect\citeauthoryear{}{}]{}
%   
%  
% \end{thebibliography}

\end{document}